\documentclass[aps, prb, twocolumn, amssymb, amsmath, showpacs, superscriptaddress]{revtex4-1}
\usepackage{bm}
\usepackage{times}
\usepackage{graphicx}
\usepackage{color}
\usepackage{dcolumn}
\usepackage[colorlinks=true, letterpaper=true, pdfstartview=FitV, linkcolor=blue, citecolor=blue, urlcolor=blue]{hyperref}
\usepackage{appendix}
\usepackage[normalem]{ulem}

\newcommand{\vect}[1]{\boldsymbol{#1}}

\usepackage{pifont}
\newcommand{\cmark}{\ding{51}}
\newcommand{\xmark}{\ding{55}}

\begin{document}

\title{Spin transport revealed by the \textit{spin quantum geometry}}
\author{Longjun Xiang}
\affiliation{College of Physics and Optoelectronic Engineering, Shenzhen University, Shenzhen 518060, China}
\author{Hao Jin}
\email{jh@szu.edu.cn}
\affiliation{College of Physics and Optoelectronic Engineering, Shenzhen University, Shenzhen 518060, China}
\author{Jian Wang}
\email{jianwang@hku.hk}
\affiliation{College of Physics and Optoelectronic Engineering, Shenzhen University, Shenzhen 518060, China}
\affiliation{Quantum Science Center of Guangdong-Hongkong-Macao Greater Bay Area (Guangdong), Shenzhen 518045, China}
\affiliation{Department of Physics, The University of Hong Kong, Pokfulam Road, Hong Kong, China}

\date{\today}

\begin{abstract}
We present the framework of \textit{spin quantum geometry},
which is fundamentally linked to the spin degree of freedom of Bloch electrons
and incorporates both the spin quantum geometric tensor (QGT) and the recently introduced Zeeman QGT,
to elucidate the spin transport.
We show that the spin and Zeeman QGTs, respectively,
provide a unified framework for revealing known spin currents, such as the intrinsic spin Hall effect,
and spin magnetization, such as the Edelstein effect, of Bloch electrons under an electric field.
In addition, we predict the linear displacement spin Hall effect,
wherein an AC electric field induces a transverse spin current in insulating systems.
Furthermore, we propose two novel nonlinear spin responses:
the nonlinear Drude spin current (NDSC) and the nonlinear Drude spin magnetization (NDSM),
both of which exhibit a quadratic dependence on the relaxation time, like the nonlinear Drude charge current,
and are governed by the \textit{spin quantum geometry}.
Finally, we evaluate the NDSC and NDSM with Dirac models of topological insulators and
find that, in the moderately dirty regime, the NDSC and NDSM can exceed
their respective nonlinear intrinsic counterparts,
which have recently garnered significant interest in spintronics.
\end{abstract}

\maketitle

\noindent \textit{\textcolor{blue}{Introduction.}}---Beyond the energy spectrum,
the quantum wave function plays a central role in determining the response behavior of Bloch electrons in solids.
In particular, as the gauge-invariant combination of the Bloch wave function, the quantum geometric tensor (QGT),
which comprises the quantum metric (QM) and the Berry curvature (BC),
offers a versatile framework for describing charge transport under external fields~\cite{NiuRMP,
QuantumGeometry, YanReview, BernivergReview, QueirozReview, HZLu2024, FuBCD, MaBCDexp, KFMBCDexp,
LuHZNRP, GaoQMD, GaoPRL2021, XiaoPRL2021, YanPRL2024, XuSY2023, Gao2023, QMNP, KDas, SuPRB, Jia2024,
XiaoCMNHE, XiangIMNHE, KTLawPRB, intrinsic3, BPT2021, WangPRL, capacitor, xiangDHE,
WYHe, JEMoore2010, MaQlight2021, MaQlight2023, Nagaosa2018NC, Yan2020, AhnPRX,
AhnNP, AMIT, DaoDao, AMIT1, BJYang, Resta, GaoY2019, XiaoCQF}.
Specifically, the $\mathcal{T}$-odd ($\mathcal{T}$, time reversal) BC
governs the intrinsic anomalous Hall effect in ferromagnets~\cite{NiuRMP};
the $\mathcal{T}$-even but $\mathcal{P}$-odd ($\mathcal{P}$, inversion)
Berry curvature dipole (BCD) drives the extrinsic nonlinear Hall effect
in noncentrosymmetric metals~\cite{FuBCD, MaBCDexp, KFMBCDexp};
the $\mathcal{P}$-odd, $\mathcal{T}$-odd, but $\mathcal{P}\mathcal{T}$-even quantum metric dipole (QMD)
causes the intrinsic nonlinear Hall effect in magnetic
materials without $\mathcal{P}$-symmetry~\cite{GaoQMD, GaoPRL2021,
XiaoPRL2021, YanPRL2024, XuSY2023, Gao2023, QMNP, KDas, SuPRB, Jia2024}.
Besides the charge transport, the responses of the spin degree of freedom in Bloch electrons under an electric field,
including the spin currents~\cite{universal, Mn3Sn, spinvorticity, XiaoCNC2024,
Todd1, Todd2, DaoDaoNC, spinBCNP, spinBCNP1, spinsym, nonlinearspin2, pureSHE, spinBC, spinBCD}
and the spin magnetization~\cite{Edelstein, XuHW2025, XiaoCspingeneration1, XiaoCspingeneration2,
Sipe2022, XiaoarXiv, naturetech, intrinsicEdel0, intrinsicEdel1, resonantspin, XuHW2021},
are also governed by the Bloch wave function.
However, a unified quantum geometric framework for spin responses is still lacking,
obscuring the comprehensive understanding of spin transport phenomena.

In this Letter, we present the framework of \textit{spin quantum geometry},
built upon the spin QGT and the recently introduced Zeeman QGT~\cite{XiangZeeman},
both of which intrinsically incorporate the spin degree of freedom, to reveal the spin transport.
We show that the spin QGT and the Zeeman QGT, respectively,
provide a unified framework for revealing spin currents and spin magnetization of Bloch electrons in response to an electric field,
as summarized in Table~\ref{tab0}. We find that \textit{all contributions} to spin currents and spin magnetization
up to second order in the electric field can be systematically expressed within this spin quantum geometric framework.
Specifically, under a DC electric field,
we show that the spin BC and Zeeman QM, respectively, govern the linear intrinsic spin currents~\cite{universal}
and spin magnetization~\cite{naturetech, intrinsicEdel0, intrinsicEdel1};
the spin QM and Zeeman BC, respectively, govern the linear extrinsic spin currents~\cite{Mn3Sn, spinvorticity, XiaoCNC2024}
and spin magnetization~\cite{Edelstein, XuHW2025};
the spin QMD and Zeeman BCD, respectively, govern the nonlinear intrinsic spin current~\cite{nonlinearspin2, pureSHE}
and spin magnetization~\cite{XiaoCspingeneration1};
the spin BCD and Zeeman QMD, respectively, govern the nonlinear extrinsic spin current~\cite{nonlinearspin2, pureSHE}
and spin magnetization~\cite{XiaoCspingeneration2},
particularly at the linear order of the relaxation time $\tau$.

Besides unifying these known spin responses,
we predict that both the spin (Zeeman) QM and spin (Zeeman) BC 
can induce a displacement spin current (spin magnetization~\cite{Sipe2022, XiaoarXiv})
in insulating materials under an AC electric field,
where the displacement spin current may exhibit a Hall component.
Moreover, we propose the nonlinear Drude spin current (NDSC) and the nonlinear Drude spin magnetization (NDSM),
which are analogous to the nonlinear Drude charge current and scale quadratically with $\tau$.
Remarkably, we find that the NDSC (NDSM) is induced by the spin QMD (Zeeman BCD),
coinciding with the spin quantum geometric origin of its nonlinear intrinsic counterpart.
Finally, through calculations based on Dirac models derived from topological insulators,
we show that the NDSC (NDSM) can exceed its nonlinear intrinsic counterpart by two (three) orders of magnitude
in the moderately dirty regime.
Our work not only uncovers three novel spin response functions with promising applications in spintronics
but also highlights the spin quantum geometric physics of Bloch electrons.

\begin{center}
\begin{table*}[t!]
\caption{\label{tab0}
The parity for the quantum metric (QM), Berry curvature (BC), quantum metric dipole (QMD)~\cite{GaoPRL2021},
and Berry curvature dipole (BCD)~\cite{FuBCD} of the spin and Zeeman QGTs under inversion $(\mathcal{P})$,
time reversal $(\mathcal{T})$, and $\mathcal{P}\mathcal{T}$ symmetries.
Here, \cmark (\xmark) indicates the even (odd) parity under the corresponding symmetry operation.
Note that the spin (Zeeman) QM and BC govern
the $\sigma^{(i)}_{ab;c}$ ($\beta^{(i)}_{ab}$) for the linear spin current (spin magnetization)
while the spin (Zeeman) QMD and BCD govern the $\sigma^{(i)}_{abd;c}$ ($\beta^{(i)}_{abc}$)
for the nonlinear spin current (spin magnetization).
Interestingly, the conventional QM can also induce the nonlinear intrinsic spin current and spin magnetization.
}
\begin{tabular}{ c| c | c | c | c | p{4.0cm} | p{4.0cm}}
\hline
\hline
Spin                   & \ QM ($g^{ab;c}_{nm}$) \
                       & \ BC ($\Omega^{ab;c}_{nm}$)  \
                       & \ QMD ($v_n^dg^{ab;c}_{nm}$) \ 
                       & \ BCD ($v_n^d\Omega^{ab;c}_{nm}$) \
                       & \ \ \text{Linear spin current}
                       & \ \ \text{Nonlinear spin current} 
\\
[0.05em]
\hline
$\mathcal{P}$ & \cmark
              & \cmark 
              & \xmark 
              & \xmark
              & \ \ $\sigma_{ab;c}^{(1)}$: spin QM~\cite{Mn3Sn, spinvorticity, XiaoCNC2024}
              & \ \ $\sigma_{abd;c}^{(2)}$: spin QMD$^\dagger$ 
\\                                                  
[0.05em]                                            
$\mathcal{T}$ & \xmark
              & \cmark
              & \cmark 
              & \xmark
              & \ \ $\sigma^{(0)}_{ab;c}$: spin BC~\cite{universal}
              & \ \ $\sigma_{abd;c}^{(1)}$: spin BCD~\cite{pureSHE}
\\
[0.05em]
$\mathcal{P}\mathcal{T}$ & \xmark
                         & \cmark 
                         & \xmark 
                         & \cmark
                         & \ \ $\sigma^{(0)}_{ab;c}(\omega)$: spin BC/QM$^\dagger$
                         & \ \ $\sigma_{abd;c}^{(0)}$: spin QMD$+$QM~\cite{nonlinearspin2, pureSHE} 
\\
[0.05em]
\hline
Zeeman         \       & QM ($\bar{g}^{ab}_{nm}$)
                       & BC ($\bar{\Omega}^{ab}_{nm}$)
                       & QMD ($v_n^c\bar{g}^{ab}_{nm}$)
                       & BCD ($v_n^c\bar{\Omega}^{ab}_{nm}$)
                       & \ \ \text{Linear spin magnetization}
                       & \ \ \text{Nonlinear spin magnetization} 
\\
[0.05em]
\hline
$\mathcal{P}$           & \xmark 
                        & \xmark 
                        & \cmark 
                        & \cmark
                        & \ \ $\beta^{(1)}_{ab}$: Zeeman BC~\cite{Edelstein, XuHW2025}
                        & \ \ $\beta_{abc}^{(2)}$: Zeeman BCD$^\dagger$
\\                                                  
[0.05em]                                            
$\mathcal{T}$           & \xmark
                        & \cmark
                        & \cmark 
                        & \xmark
                        & \ \ $\beta^{(0)}_{ab}$: Zeeman QM~\cite{naturetech, intrinsicEdel0, intrinsicEdel1}
                        & \ \ $\beta_{abc}^{(1)}$: Zeeman QMD~\cite{XiaoCspingeneration2}
\\
[0.05em]
$\mathcal{P}\mathcal{T}$ & \cmark 
                         & \xmark
                         & \cmark
                         & \xmark
                         & \ \ $\beta^{(0)}_{ab}(\omega)$: Zeeman BC/QM~\cite{Sipe2022, XiaoarXiv}
                         & \ \ $\beta_{abc}^{(0)}$: Zeeman BCD$+$QM~\cite{XiaoCspingeneration1}
\\
[0.05em]
\hline
\hline
\end{tabular}
\end{table*}
\end{center}

\bigskip
\noindent \textit{\textcolor{blue}{Spin quantum geometry.}}---In solid-state physics,
the QGT $T^{ab}_{nm}$ is defined by~\cite{AhnNP, note} ($e=\hbar=1$)
\begin{align}
T^{ab}_{nm}=r^a_{nm}r^b_{mn},
\label{QGT}
\end{align}
where $a, b \in \{x, y, z\}$ and $r_{nm}^a \equiv \langle u_n|\hat{r}^a|u_m\rangle$
with $n \neq m$ is the interband Berry connection.
Here, $|u_m\rangle$ is the cell-periodic Bloch wave function
and $\hat{r}^a=i\partial/\partial k_a=i\partial_a$ is the position operator in momentum space,
with $k_a$ being the crystal momentum.
Eq.~(\ref{QGT}) is gauge-invariant but generally complex, which can be written as $T^{ab}_{nm}=g^{ab}_{nm}-\frac{i}{2}\Omega^{ab}_{nm}$,
where $\Omega^{ab}_{nm} \equiv -2\text{Im}[r^a_{nm}r^b_{mn}]$
is the (local) BC~\cite{NiuRMP},
and $g^{ab}_{nm} \equiv \text{Re}[r^a_{nm}r^b_{mn}]$ is the (local) QM,
which has recently attracted significant interest~\cite{GaoQMD,
GaoPRL2021, XiaoPRL2021, YanPRL2024, XuSY2023, Gao2023, QMNP, KDas, SuPRB, Jia2024}.

The conventional QGT can be extended to incorporate the spin degree of freedom
representing by the Pauli matrix $\hat{\sigma}^c$ with $c \in \{x, y, z\}$.
To that purpose, we rewrite Eq.~(\ref{QGT}) into $T_{nm}^{ab}=v^a_{nm}v^b_{mn}/\epsilon_{nm}^2$
by using~\cite{AhnNP} $r^a_{nm}=-iv^a_{nm}/\epsilon_{nm}$ for $n \neq m$,
where $\epsilon_{nm}=\epsilon_n-\epsilon_m$ is the band energy difference
and $v^a_{nm}=\langle u_n|\hat{v}^a|u_m\rangle$ is the interband matrix element
of the velocity operator $\hat{v}^a$.
Then by replacing $v^a_{nm}$ with $v^{a;c}_{nm}=\langle u_n|\hat{v}^{a;c}|u_m\rangle$,
where ${\hat v}^{a;c}=\frac{1}{4}\{\hat{v}^a,\hat{\sigma}^c\}$ is the spin-current operator,
we obtain the spin QGT
\begin{align}
T_{nm}^{ab;c}
\equiv
v^{a;c}_{nm} v^b_{mn}/\epsilon_{nm}^2,
\label{spinQGT}
\end{align}
which is gauge-invariant but generally complex.
As a result, by defining $T_{nm}^{ab;c}=g^{ab;c}_{nm}-\frac{i}{2}\Omega^{ab;c}_{nm}$,
we identify $\Omega_{nm}^{ab;c}\equiv-2\text{Im}[v^{a;c}_{nm}v^b_{mn}]/\epsilon_{nm}^2$
as the (local) spin BC, which has been established in understanding the intrinsic spin Hall effect~\cite{universal}.
Correspondingly, $g_{nm}^{ab;c} \equiv \text{Re}[v^{a;c}_{nm}v^b_{mn}]/\epsilon_{nm}^2$
is interpreted as the (local) \textit{spin QM}---the quantum geometric counterpart of the spin BC.
Although similar expressions have appeared in earlier works~\cite{Todd1, Todd2, XiaoCNC2024},
we wish to remark that the spin QM has not been explicitly discussed from a quantum geometric perspective.
As we will show below, the spin QM will play an equally important role with the spin BC
in unveiling the linear and nonlinear spin currents of Bloch electrons.

Recently, by evaluating the quantum distance of the Hilbert space spanned by the spinor Bloch states,
along with Eq.~(\ref{QGT}),
the gauge-invariant Zeeman QGT was proposed~\cite{XiangZeeman}, which is defined as
\begin{align}
\bar{T}^{ab}_{nm} = r^a_{nm} \sigma^b_{mn}.
\label{ZeemanQGT}
\end{align}
Here $\sigma^b_{mn} \equiv \langle u_m|\hat{\sigma}^b|u_n\rangle$ explicitly incorporate the spin degree of freedom.
Similarly, we define $\bar{T}^{ab}_{nm}=\bar{g}^{ab}_{nm}-\frac{i}{2}\bar{\Omega}^{ab}_{nm}$,
where $\bar{\Omega}^{ab}_{nm} \equiv -2\text{Im}[r^a_{nm}\sigma^b_{mn}]$
and $\bar{g}^{ab}_{nm} \equiv \text{Re}[r^a_{nm}\sigma^b_{mn}]$
give rise to the (local) Zeeman BC and the (local) Zeeman QM, respectively.
These quantities have been shown to play a key role in generating gyrotropic magnetic current~\cite{XiangZeeman, JEMoore2016}.
As we will demonstrate below, the Zeeman QGT can be further used to reveal
the linear and nonlinear spin magnetization of Bloch electrons.

We remark that Eqs.~(\ref{spinQGT})-(\ref{ZeemanQGT})
can be broadly identified as the \textit{spin quantum geometry},
particularly due to their explicit incorporation of the spin degree of freedom
and their explicit dependence on spin-orbit coupling~\cite{SOC}.
In addition, we note that the QMs and BCs of Eqs.~(\ref{spinQGT})-(\ref{ZeemanQGT})
usually show different parities with that of Eq.~(\ref{QGT}) under the same symmetry operation.
\textcolor{blue}{For example, using $\mathcal{T}r^a_{nm}=r^a_{mn}$, $\mathcal{T}\sigma^a_{nm}=-\sigma^a_{mn}$,
and $\mathcal{T}v^{a;c}_{nm}=v^{a;c}_{mn}$,
we find that the spin BC $\Omega^{ab;c}_{nm}$ and the Zeeman BC $\bar{\Omega}^{ab}_{nm}$ are $\mathcal{T}$-even
while the conventional BC $\Omega^{ab}_{nm}$ is $\mathcal{T}$-odd.}
For convenience, in Table~\ref{tab0},
we list the parity of the QM, BC, QMD~\cite{GaoPRL2021}, and BCD~\cite{FuBCD} of the spin and Zeeman QGTs
under $\mathcal{P}$, $\mathcal{T}$, and $\mathcal{P}\mathcal{T}$ symmetries.
After elaborating the \textit{spin quantum geometry},
we proceed to demonstrate how it can be used
to reveal the spin transport of Bloch electrons under an electric field,
following the same spirt that the charge transport is unveiled by the conventional QGT.
Particularly, by using the spin QGT and Zeeman QGT, respectively,
we unify that the known spin currents and spin magnetization of Bloch electrons
up to the second order of the electric field.
Furthermore, we uncover three novel spin quantum geometric response functions.

\bigskip
\noindent \textit{\textcolor{blue}{Linear spin current and spin magnetization.}}---The spin current,
denoted as $J_{a;c}$ (with $c$ indicating the spin polarization direction),
and the spin magnetization, denoted as $M_a$, induced linearly by an electric field $E_b$,
can be generally written as $J_{a;c}^{(1)} = \sigma_{ab;c} E_b$ and $M_a^{(1)} = \beta_{ab} E_b$, respectively.
Here $\sigma_{ab;c}$ and $\beta_{ab}$ are the corresponding response tensors.
According to their dependence on the relaxation time $\tau$,
we decompose them as~\cite{GaoY2019}
$\sigma_{ab;c} = \sigma^{(1)}_{ab;c} + \sigma^{(0)}_{ab;c}$
and
$\beta_{ab} = \beta^{(1)}_{ab} + \beta^{(0)}_{ab}$,
where $\sigma_{ab;c}^{(1)}$ and $\beta^{(1)}_{ab}$ linearly scaling with $\tau$ represent extrinsic contributions, while
$\beta^{(0)}_{ab}$ and $\sigma^{(0)}_{ab;c}$ free of $\tau$ are intrinsic components.
Throughout this work, we adopt the Einstein summation convention for the repeated indices associated with the electric field.

Using the density matrix formalism~\cite{Sipe1995, Sipe2000, Jia2024},
the response tensor $\sigma_{ab;c}$ for the spin current 
and the response tensor $\beta_{ab}$ for the spin magnetization
can be derived in a straightforward way,
as detailed in the Supplementary Material~\cite{sup}.
In the DC limit, for the extrinsic contributions, we find~\cite{sup}
\begin{align}
\sigma_{ab;c}^{(1)} &= - 2 \tau \sum_{nm} \int_k f_n \epsilon_{nm} g^{ab;c}_{nm}, \label{sigma1} \\
\beta_{ab}^{(1)} &= - \frac{\tau}{2} \sum_{nm} \int_k  f_n \bar{\Omega}^{ba}_{nm},
\label{beta1}
\end{align}
where $\int_k=\int d\vect{k}/(2\pi)^d$ and
$f_n$ is the equilibrium Fermi-Dirac distribution function.
We remark that Eq.~(\ref{sigma1}), governed by the $\mathcal{T}$-odd spin QM~\cite{foot0} $g^{ab;c}_{nm}$,
describes the magnetic spin Hall effect,
as observed in the non-collinear antiferromagnet Mn$_3$Sn~\cite{Mn3Sn, spinvorticity}
and recently in magnetic van der Waals heterostructure Fe$_3$GeTe$_2$/MoTe$_2$~\cite{XiaoCNC2024}.
In contrast, Eq.~(\ref{beta1}), governed by the $\mathcal{T}$-even Zeeman BC $\bar{\Omega}^{ba}_{nm}$,
captures the well-known Edelstein effect~\cite{XuHW2025, Edelstein}.
For the intrinsic contributions, we find~\cite{sup}
\begin{align}
\sigma_{ab;c}^{(0)} = \sum_{nm} \int_k f_n \Omega^{ab;c}_{nm},
\quad
\beta_{ab}^{(0)} = \sum_{nm} \int_k f_n \frac{\bar{g}^{ba}_{nm}}{\epsilon_{nm}},
\label{sigma0}
\end{align}
where the $\mathcal{T}$-even spin BC $\Omega^{ab;c}_{nm}$
induces the intrinsic spin Hall effect proposed by Sinova \textit{et al.}~\cite{universal};
the $\mathcal{T}$-odd Zeeman QM $\bar{g}^{ba}_{nm}$ causes the intrinsic spin magnetization
in magnetic insulators~\cite{naturetech, intrinsicEdel0, intrinsicEdel1}.

Beyond the DC limit, the intrinsic spin current and spin magnetization
can further include the frequency-dependent contributions~\cite{sup}
\begin{align}
\sigma^{(0)}_{ab;c}(\omega)
&=
\sum_{nm} \int_k f_n (\lambda_1^{nm} \Omega^{ab;c}_{nm}+2\lambda^{nm}_2 g^{ab;c}_{nm}),
\label{sigma0w}
\\
\beta^{(0)}_{ab} (\omega)
&=
\sum_{nm} \int_k f_n
\left(
\lambda^{nm}_1 \frac{\bar{g}^{ba}_{nm}}{\epsilon_{nm}} + \frac{\lambda^{nm}_2}{2} \frac{\bar{\Omega}^{ba}_{nm}}{\epsilon_{nm}} \right),
\label{beta0w}
\end{align}
where $\lambda_1^{nm}=-\omega^2\cos( \omega t)/(\omega^2-\epsilon_{nm}^2)$
and $\lambda_2^{nm}=\epsilon_{nm}\omega\sin(\omega t)/(\omega^2-\epsilon_{nm}^2)$.
In stark contrast with Eq.~(\ref{sigma0}), under an AC electric field,
both the spin BC and spin QM can contribute to the intrinsic spin current $\sigma^{(0)}_{ab;c}$
and both the Zeeman BC and Zeeman QM to the intrinsic spin magnetization $\beta^{(0)}_{ab}$.
\textcolor{blue}{We wish to mention that Eqs.~(\ref{sigma0w})-(\ref{beta0w})
are derived analogous to the displacement charge current~\cite{xiangDHE}
and hence, we dub them the displacement spin current and the displacement spin magnetization~\cite{displacement}, respectively.
While the intrinsic displacement spin magnetization has been recently discussed~\cite{Sipe2022, XiaoarXiv},
the intrinsic displacement spin current has not yet been proposed.
Notably, $\sigma_{ab;c}^{(0)}(\omega)$ with $a \neq b$ gives an intrinsic displacement spin Hall current,
which we explore through both model and first-principles calculations in the Supplementary Material~\cite{sup}.}

As expected, we find that the linear spin current and spin magnetization, respectively,
can be systematically described by the spin and Zeeman QGT, as summarized in Table~\ref{tab0}.
To close this section, we remark that Eqs.~(\ref{sigma1})-(\ref{beta1})
and Eqs.~(\ref{sigma0})-(\ref{beta0w}) feature the Fermi-surface property~\cite{Fermisurf}
and the Fermi-sea property, respectively. This indicates that the former can only appear
in metallic materials, while the latter can occur in both metallic and insulating materials.

\bigskip
\noindent \textit{\textcolor{blue}{Nonlinear spin current and spin magnetization.}}---Beyond
linear spin responses, the \textit{spin quantum geometry}
can also reveal the nonlinear spin responses, including the spin currents and spin magnetization.
As an illustration, we focus on the DC limit.
In the second order of the electric field, we define the spin current and spin magnetization as
$J_{a;c}^{(2)}=\sigma_{abd;c}E_bE_d$
and
$M_a^{(2)}=\beta_{abc}E_bE_c$, respectively.
Further, the second-order response tensors $\sigma_{abd;c}$ and $\beta_{abc}$ can be decomposed as~\cite{GaoY2019}
$\sigma_{abd;c} = \sum_{i=0}^{i=2} \sigma_{abd;c}^{(i)}$
and
$\beta_{abc} = \sum_{i=0}^{i=2} \beta_{abc}^{(i)}$,
where $\sigma_{abd;c}^{(i)}\propto \tau^i$ and $\beta_{abc}^{(i)}\propto \tau^i$.
This decomposition exhibits the classification of their extrinsic (depending on $\tau$)
and intrinsic (free of $\tau$) nature and further facilitate the identification
of their spin quantum geometric origin,
the same as the linear spin transport discussed above.

We begin by examining the extrinsic contributions. At the second order of $\tau$, we find~\cite{sup}
\begin{align}
\sigma_{abd;c}^{(2)}
&=
-
\tau^2\sum_{nm} \int_k
\epsilon_{nm} g^{ab;c}_{nm}
\partial_d f_n,
\label{sigma22}
\\
\beta_{abc}^{(2)} &= -\frac{\tau^2}{4} \sum_{nm} \int_k \bar{\Omega}^{ba}_{nm} \partial_c f_n. \label{beta22}
\end{align}
Due to $\partial_a f_n = v_n^a \partial f_n/ \partial \epsilon_n$,
it is evident that Eq.~(\ref{sigma22}) and Eq.~(\ref{beta22}),
respectively, arise from the spin QMD $v_n^d g^{ab;c}_{nm}$
and the Zeeman BCD $v_n^c \bar{\Omega}^{ba}_{mn}$,
which are defined similarly to the conventional QMD~\cite{GaoPRL2021}.
Note that Eqs.~(\ref{sigma22})-(\ref{beta22}) are derived using
the density matrix element used to develop the nonlinear Drude charge current~\cite{Jia2024},
as indicated by their quadratic dependence on $\tau$. 
Hence, the nonlinear spin current from $\sigma_{abd;c}^{(2)}$ and the nonlinear spin magnetization from $\beta_{abc}^{(2)}$,
respectively, are dubbed the nonlinear Drude spin current (NDSC)
and nonlinear Drude spin magnetization (NDSM).
At the linear order of $\tau$, we obtain~\cite{sup}
\begin{align}
\sigma^{(1)}_{abd;c}
&=
\tau \sum_{nm} \int_k
\Omega^{ab;c}_{nm} \partial_d f_n,
\label{sigma21}
\\
\beta_{abc}^{(1)} &=
\tau \sum_{nm} \int_k \frac{\bar{g}^{ba}_{nm}}{\epsilon_{nm}} \partial_c f_n,
\label{beta21}
\end{align}
where Eq.~(\ref{sigma21}), governed by the spin BCD $v_n^d \bar{\Omega}^{ab;c}_{nm}$,
has recently appeared in Ref.~[\onlinecite{pureSHE}]; Eq.~(\ref{beta21}),
governed by the Zeeman QMD $v_n^c \bar{g}^{ba}_{nm}$,
has been derived using the semiclassical theory to achieve the spin generation
in centrosymmetric materials~\cite{XiaoCspingeneration2}.
\textcolor{blue}{We wish to mention that the quantity $2 \sum_{m} \bar{g}^{ba}_{nm}/\epsilon_{nm}$
was identified as \textit{anomalous spin polarizability} in Ref.~[\onlinecite{XiaoCspingeneration2}].}

Finally, the response tensors for the intrinsic nonlinear spin current and spin magnetization,
respectively, are given by~\cite{sup}
\begin{align}
\sigma_{abd;c}^{(0)}
&=
\sum_{nm} \int_k
\left(
\frac{
\mathcal{Q}^{abd;c}_{nm}
}{\epsilon_{nm}^2}
f_n
+
\frac{2g^{ad;c}_{nm} }{\epsilon_{nm}}
\partial_b f_n
\right)
,
\label{sigma20}
\\
\beta_{abc}^{(0)}
&=
\sum_{nm}
\int_k
\left(
\frac{
\mathcal{Z}_{nm}^{abc}
}{\epsilon_{nm}^3}
f_n
+
\frac{\bar{\Omega}^{ca}_{nm}}{2\epsilon_{nm}^2}
\partial_b f_n
\right).
\label{beta20}
\end{align}
Here $\mathcal{Q}_{nm}^{abd;c} = 2\Delta^d_{mn}g^{ab;c}_{nm}+\bar{\Delta}^{a;c}_{mn} g^{bd}_{nm}$,
where $\Delta^d_{mn}=\partial_d\epsilon_{mn}$ and $\bar{\Delta}^{a;c}_{mn}=v^{a;c}_{mm}-v^{a;c}_{nn}$;
$\mathcal{Z}_{nm}^{abc}=\Delta^c_{mn}\bar{\Omega}^{ba}_{nm}/2+\bar{\Delta}^a_{mn} \epsilon_{nm} g^{bc}_{nm}$,
where $\bar{\Delta}^a_{mn}=(\sigma^a_{mm}-\sigma^a_{nn})/2$.
The same as the NDSC and NDSM, we note that Eq.~(\ref{sigma20}) and Eq.~(\ref{beta20})
can also arise from the spin QMD ($\Delta^d_{mn}g^{ab;c}_{nm}$ or $v_n^dg^{ad;c}_{nm}$) and 
the Zeeman BCD ($\Delta^c_{mn}\bar{\Omega}^{ba}_{nm}$ or $v_n^b\bar{\Omega}^{ca}_{nm}$), respectively.
Besides, we note that both Eq.~(\ref{sigma20}) and Eq.~(\ref{beta20})
can originate from the conventional QM ($g^{ab}_{nm}$)~\cite{XiaoCspingeneration1, nonlinearspin2}.
We remark that similar expressions to Eq.~(\ref{sigma20}) have been derived
to achieve the pure intrinsic spin Hall current~\cite{pureSHE} and
to realize magnetic-field-free perpendicular magnetization switching~\cite{nonlinearspin2},
where $2\sum_{m} g^{ad;c}_{nm}/\epsilon_{nm}$ is identified as the \textit{spin Berry connection polarizability}.
\textcolor{blue}{Likewise, similar expressions to Eq.~(\ref{beta20}) have been derived
to realize the spin generation in centrosymmetric magnets~\cite{XiaoCspingeneration1},
where $2\sum_{m}\bar{\Omega}^{ca}_{nm}/\epsilon_{nm}^2$
is interpreted as the \textit{$h$-space Berry connection polarizability}.}

After revealing the spin quantum geometric origin of the nonlinear spin current and spin magnetization,
as summarized in Table.~\ref{tab0}, we remark that
Eqs.~(\ref{sigma22})-(\ref{beta21}) account solely for Fermi-surface contributions, 
while Eqs.~(\ref{sigma20})-(\ref{beta20}) contain both the Fermi-surface and Fermi-sea contributions.
In addition, we note that Eqs.~(\ref{sigma20})-(\ref{beta20}) neglect four-band and three-band processes, respectively,
whose influence are generally considered negligible~\cite{Threeband1, Threeband2, Threeband3, Threeband4}.
To close this section, we emphasize that the NDSC described by Eq.~(\ref{sigma22})
and the NDSM given by Eq.~(\ref{beta22}) have not been previously explored
from the perspective of quantum geometry, as will be immediately discussed below.

\begin{figure}[t!]
\includegraphics[width=0.95\columnwidth]{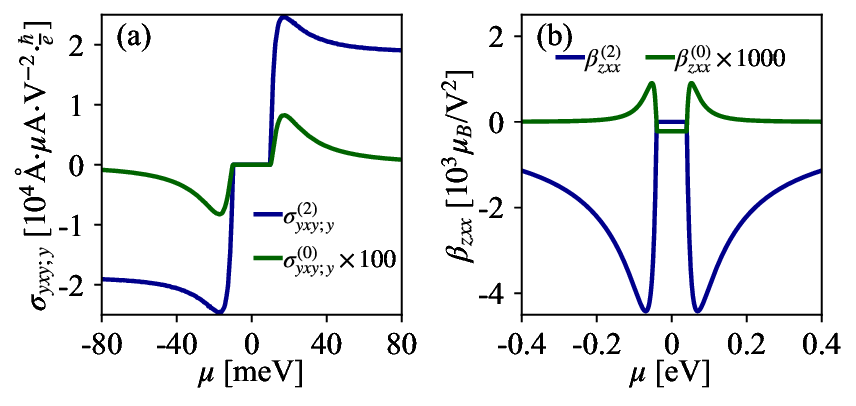}
\caption{
(a) Nonlinear Drude spin Hall conductivity $\sigma^{(2)}_{yxy;y}$ versus
nonlinear intrinsic spin Hall conductivity $\sigma^{(0)}_{yxy;y}$.
Parameters: $\tau=0.5 \mathrm{ps}$~\cite{KTLawPRB}, $\beta=10 \mathrm{meV}$, $v_x=v_y=10^5 \mathrm{m/s}$, and $\alpha=0.1 v_x$~\cite{FuBCD}.
(b) $\beta^{(2)}_{zxx}$ for NDSM versus its intrinsic counterpart $\beta^{(0)}_{zxx}$.
Parameters: $\tau=0.5 \mathrm{ps}$~\cite{KTLawPRB} and $\Delta=40 \mathrm{meV}$.
}
\label{FIG1}
\end{figure}

\bigskip
\noindent \textit{\textcolor{blue}{Nonlinear Drude spin current.}}---We note that $\sigma_{abd;c}^{(2)}$
for NDSC decided by the spin QMD is $\mathcal{T}$-even but $\mathcal{P}$-odd, see Table~\ref{tab0},
where $\sigma_{abd;c}^{(2)}$ with $a \notin \{b, d\}$ gives rise to the nonlinear Drude spin Hall current.
Guided by symmetry, we investigate this effect using the
Dirac model~\cite{FuBCD} $H_{s\Lambda}=v_xk_x\hat{\sigma}^y-sv_yk_y\hat{\sigma}^x+\beta\hat{\sigma}^z+s\alpha k_y$,
which effectively describes a pair of massive Dirac cones located at the momenta $\pm \Lambda$
in the surface Brillouin zone of topological crystalline insulators such as SnTe~\cite{SnTe}, 
and has been widely used to study the nonlinear Hall effect induced by the conventional BCD~\cite{FuBCD}.
Here, $\hat{\sigma}^a$ stands for pseudospin,
$s=\pm 1$, $v_x$ and $v_y$ are Fermi velocities, $\beta$ is the gap parameter,
and $\alpha$ is the tilt parameter. For this model, the relevant component of the spin QM
is given by $g_{\pm\mp}^{yx;y}=s v_x \alpha (v_y^2k_y^2+\beta^2)/[8(v_x^2k_x^2+v_y^2k_y^2+\beta^2)^2]$,
which is even in $k_x$ and $k_y$.
Considering its dipole along the tilt direction,
using Eq.~(\ref{sigma22}) we numerically evaluate the nonlinear Drude spin Hall conductivity $\sigma_{yxy;y}^{(2)}$
at zero temperature for various chemical potentials $\mu$
by chosing $\tau=0.5 \mathrm{ps}$~\cite{KTLawPRB}.
The result is shown in Fig.~\ref{FIG1}(a),
where we observe a large peak near the gap, accompanied by a sign change as a function of $\mu$.
For comparison, we also show the nonlinear intrinsic spin Hall conductivity $\sigma_{yxy;y}^{(0)}$ in Fig.~\ref{FIG1}(a).
We find that $\sigma_{yxy;y}^{(2)}$ is two orders larger than $\sigma_{yxy;y}^{(0)}$.
In addition, the in-gap contribution of $\sigma_{yxy;y}^{(0)}$ due to the Fermi-sea contribution is negligible.

Beyond the model calculations,
we note that both $\sigma_{abd;c}^{(2)}$ and the Fermi-surface part of $\sigma_{abd;c}^{(0)}$
are contributed by the spin QMD.
As a result, by a simple dimension analysis, we find $\sigma_{abd;c}^{(2)}/\sigma_{abd;c}^{(0)} \sim \tau^2 \mu^2/\hbar^2$.
In a moderately dirty simple ($\tau\sim 10^{-13} \mathrm{s}$) with light doping ($\mu\sim 10^{-2} \mathrm{eV}$),
we find that $\sigma_{abd;c}^{(2)}/\sigma_{abd;c}^{(0)} \sim 2$.
In this regime, the nonlinear Drude spin (Hall) current will
compete with its nonlinear intrinsic counterpart~\cite{XiaoCspingeneration1}.

\bigskip
\noindent \textit{\textcolor{blue}{Nonlinear Drude spin magnetization.}}---On the other hand,
we note that $\beta_{abc}^{(2)}$ for NDSM decided by the Zeeman BCD is $\mathcal{T}$-odd but $\mathcal{P}$-even,
see Table~\ref{tab0}.
Therefore, we can employ the massive Dirac model~\cite{GaoQMD} $H=v(k_x\hat{\sigma}^x+k_y\hat{\sigma}^y)+\Delta\hat{\sigma}^z$,
which breaks $\mathcal{T}$ symmetry and can be realized in the surface of topological insulator~\cite{Kane,SCZhang},
to investigate the NDSM.
Here, $v$ is the Fermi velocity and $\Delta$ is the gap parameter.
For this model, the contributing Zeeman BCs are given by
$\bar{\Omega}^{xz}_{\pm\mp}=\mp v^2\Delta k_x/(v^2k^2+\Delta^2)^{3/2}$
and
$\bar{\Omega}^{yz}_{\pm\mp}=\mp v^2\Delta k_y/(v^2k^2+\Delta^2)^{3/2}$,
where $\pm$ stands for the upper (lower) band and $k^2=k_x^2+k_y^2$.
As a representative, we consider the NDSM given by $v_{\pm}^x\bar{\Omega}_{\pm\mp}^{xz}$.
At zero temperature, after integrating $v^x_{\pm}\bar{\Omega}^{xz}_{\pm\mp}$ in terms of Eq.~(\ref{sigma22}),
we find~\cite{sup}
$\beta^{(2)}_{zxx}
=
\frac{\mu_Be^2}{\hbar^2}
\frac{\tau^2}{16\pi}
\frac{\Delta(\mu^2-\Delta^2)}{|\mu|^3}
$.
Here $e$ and $\hbar$ are restored by dimension analysis,
$\mu \notin (-\Delta, \Delta)$ is the chemical potential,
and $\mu_B$ is the Bohr magneton.

Note that the NDSM also appears at the same time with its intrinsic counterpart.
Using Eq.~(\ref{sigma20}), for the same model we find~\cite{sup}
$\beta^{(0)}_{zxx}=
\mu_B e^2 \left[\frac{3(1+\pi)\Delta^3-5\pi\Delta|\mu|^2}{960\pi^2|\mu|^5}+\frac{\Delta(|\mu|^2-\Delta^2)}{960\pi^2|\mu|^5}\right]$
when $\mu \notin (-\Delta, \Delta)$,
which includes both the Fermi-sea and Fermi-surface contributions;
while $\beta^{(0)}_{zyy} = \mu_B e^2 \frac{3(1-2\pi)\Delta^3}{960\pi^2|\mu|^5}$ when $\mu \in [-\Delta, \Delta]$,
which only contains the Fermi-sea contribution.
In Fig.~\ref{FIG1}(b), by chosing $\tau=0.5 \mathrm{ps}$~\cite{KTLawPRB},
we present the dependence of $\beta^{(2)}_{zxx}$ and $\beta^{(0)}_{zxx}$ on the chemical potential $\mu$
and we find that $\beta_{zxx}^{(2)}$ is three orders larger than $\beta_{zxx}^{(0)}$,
which means that the NDSM dominates over its intrinsic counterpart in this system.
Similar to the NDSC, in a moderately dirty simple ($\tau\sim 10^{-13} s$) with light doping ($\mu\sim 10^{-2} \mathrm{eV}$),
by a similar analysis we arrive at $\beta_{abc}^{(2)}/\beta_{abc}^{(0)} \sim 2$,
which implies that the NDSM will compete with its intrinsic counterpart~\cite{nonlinearspin2}.

\bigskip
\noindent \textit{\textcolor{blue}{Outlooks.}}---We
enclose by noting that Eqs.~(\ref{beta1})-(\ref{sigma20}) for spin transport,
when combined with first-principles simulations,
can explain the experimental observations
and guide the design of the spintronic quantum materials.
In addition, Eqs.~(\ref{spinQGT})-(\ref{ZeemanQGT}) can be utilized to
reveal the resonant spin current~\cite{DaoDaoNC} and spin magnetization~\cite{resonantspin, XuHW2021} of Bloch electrons
under light illumination, although this work focuses on the non-resonant spin current and spin magnetization.
Furthermore, based on the established spin quantum geometric framework,
we can anticipate a variety of spintronic applications.
First, owing to their common spin quantum geometric origin,
the NDSC can also facilitate the field-free switching of perpendicular magnetization
like the nonlinear intrinsic spin Hall current~\cite{nonlinearspin2},
while the NDSM can generate the spin polarization in centrosymmetric magnets
like the nonlinear intrinsic spin magnetization~\cite{XiaoCspingeneration1}.
Second, the linear displacement spin Hall current,
as discussed in detail in the Supplementary Material~\cite{sup},
may present a promising mechanism for the ultrafast spintronics~\cite{fast1, fast2, fast3, fast4}.
Finally, the same as the conventional QMD,
we note that the spin BCD is $\mathcal{P}$-odd, $\mathcal{T}$-odd, but $\mathcal{PT}$-even (see Table.~\ref{tab0})
and hence the spin current induced by the spin BCD [Eq. (\ref{sigma21})]
can also be employed to detect the reversal of the N\'eel vector in antiferromagnetic spintronics,
similar to the intrinsic nonlinear anomalous Hall current~\cite{GaoPRL2021, XiaoPRL2021} induced by the conventional QMD.

\bigskip
We acknowledge support from the National Natural Science Foundation of China (Grants No. 12034014 and No. 12404059).
H. J. thanks support from the Guangdong Basic and Applied Basic Research Foundation (Grant No. 2022A1515012006).


\begin{thebibliography}{00}
\bibitem{note}
The QGT defined by Eq.~(\ref{QGT})
particularly without summing over all possible intermediate states $|u_m\rangle$
is a local geometric quantity, which is also referred as Hermitian metric in Ref.~[\onlinecite{AhnNP}]
and is identical to the Fubini–Study metric in two-band systems.
\bibitem{foot0}
In addition to $g^{ab;c}_{nm}$, we mention that $\sigma_{ab;c}^{(1)}$ may also be contributed by
$v^{ab;c}_{nn}=\langle u_n|\{\partial_b\hat{v}^a,\hat{\sigma}^c\}|u_n\rangle/2
=\langle u_n|\{\partial_a\partial_bH, \hat{\sigma}^c\}|u_n\rangle/2$,
where $H$ is the system Hamiltonian.
However, when $H$ is linear or quadratic in $k_a$,
$v^{ab;c}_{nn}$ does not contribute to the spin Hall current,
which is usually the quantity of interest, and can thus be neglected.
\bibitem{Fermisurf}
Eqs.~(\ref{sigma1})-(\ref{beta1}) can be recast into the Fermi-surface form,
respectively, by using $\sum_{m}\bar{\Omega}^{ba}_{nm}=\partial_b \sigma^a_{nn}$
and $\sum_{m}2\epsilon_{nm}g^{ab;c}_{nm}=\partial_b v_{nn}^{a;c}$~\cite{foot0}.
\end{thebibliography}
\end{document}